\pdfoutput=1

\documentclass[final,times,5p,sort&compress]{elsarticle}

\usepackage{graphicx}
\usepackage{amssymb,amsmath}
\usepackage{txfonts}
\usepackage{natbib}
\usepackage{hyperref}

\journal{International Journal of Non-Linear Mechanics}

\begin{document}

\begin{frontmatter}
\title{Determining the Newton-Raphson basins of attraction in the electromagnetic \\
Copenhagen problem}

\author[]{Euaggelos E. Zotos\corref{cor}}
\ead{evzotos@physics.auth.gr}

\cortext[cor]{Corresponding author}

\address{Department of Physics, School of Science, \\
Aristotle University of Thessaloniki, \\
GR-541 24, Thessaloniki, \\
Greece}

\begin{abstract}
The Copenhagen problem where the primaries of equal masses are magnetic dipoles is used in order to determine the Newton-Raphson basins of attraction associated with the equilibrium points. The parametric variation of the position as well as of the stability of the Lagrange points are monitored when the value of the ratio $\lambda$ of the magnetic moments varies in predefined intervals. The regions on the configuration $(x,y)$ plane occupied by the basins of convergence are revealed using the multivariate version of the Newton-Raphson iterative scheme. The correlations between the basins of attraction of the libration points and the corresponding number of iterations needed for obtaining the desired accuracy are also illustrated. We perform a thorough and systematic numerical investigation by demonstrating how the dynamical quantity $\lambda$ influences the shape, the geometry and also the degree of fractality of the attracting domains. Our numerical results strongly indicate that the ratio $\lambda$ is indeed a very influential parameter in the electromagnetic binary system.
\end{abstract}


\end{frontmatter}

\section{Introduction}
\label{intro}

There is no doubt that one of the most intriguing topics in nonlinear dynamics is the circular restricted three-body problem \cite{S67}. A special case of this problem is the so-called ``Copenhagen problem" where the two primary bodies, which rotate with constant angular velocity around their common center of gravity, have equal masses. The third body, which is usually refereed as test particle, moves under the resultant Newtonian gravitational field of the primaries.

In \cite{KGK12} (hereafter Paper I) a very interesting version of the Copenhagen problem had been investigated. This version, known as the electromagnetic Copenhagen problem, considers the case where the two primary bodies are also magnetic dipoles, while the test particle is a small charge which moves in the electromagnetic field created by the rotating dipoles. The orbital dynamics of a charged particle moving inside the field of a electromagnetic binary system has also been investigated in previous works (e.g., \cite{K89,KM87}).

The dynamical system of electromagnetic binary system has numerous applications in several fields of physics. Since the 1990, where the exploration of exosolar planetary systems has been initiated, we have identified, until today (January 2017), more than 2600 new planetary systems (\url{http://exoplanets.eu}). Our observations indicate that a large portion of these systems are in fact two-body binary systems, where the masses of the two primary bodies (planets) are almost equal. Furthermore, the electromagnetic Copenhagen problem has also applications in astrophysics, if we take into consideration that the existence of binary systems of magnetic stars has already been confirmed.

In Paper I the authors studied the equilibrium positions of the test particle and their parametric variations, as well as the basins of attraction for two numerical methods. In particular the Newton-Raphson and the modified Broyden's methods had been used in order to determine the corresponding attracting domains. It was found that the Newton-Raphson method is much faster than the modified Broyden's method. Additionally, in the modified Broyden's method a substantial number of non-converging points exist, while on the other hand in the Newton-Raphson scheme the same type of initial conditions are very limited. In the present paper we shall expand the investigation initiated in Paper I. Taking into account that the Newton-Raphson method had been proved both faster and more accurate we shall use it in order to perform a thorough and systematic numerical exploration of the configuration plane. Our aim will be to locate the several basins of convergence and reveal how the ratio of the magnetic moments influence their structures.

At this point it should be emphasized that over the years many analytical numerical methods for solving systems of nonlinear algebraic equations have been proposed (see e.g., \cite{DB98,VI86}).

The determination of the basins of attraction for the equilibrium points (which act as attractors) using an iterative scheme is an issue of paramount importance in dynamical systems. In other words, the sets of initial conditions on the configuration $(x,y)$ plane which lead to a specific equilibrium point (attractor) define the several attraction regions (known also as basins of convergence or attracting domains). Over the last years, the basins of attraction in several types of dynamical systems have been numerically investigated. In \cite{D10} the Newton-Raphson iterative method was used in order to explore the basins of attraction in the Hill's problem with oblateness and radiation pressure. In the same vein, the multivariate version of the same iterative scheme has been used to unveil the basins of convergence in the restricted three-body problem with oblateness and radiation pressure (e.g., \cite{Z16}), the photogravitational Copenhagen problem (e.g., \cite{K08}), the four-body problem (e.g., \cite{BP11,KK14,Z17}), the ring problem of $N + 1$ bodies (e.g., \cite{CK07,GKK09}), or even the restricted 2+2 body problem (e.g., \cite{CK13}).

The present paper is organized as follows: In Section \ref{mod} we describe the basic properties of the considered mathematical model. In section \ref{lgevol} the parametric evolution of the position of the equilibrium points is investigated as the value of ratio $\lambda$ of the magnetic moments varies in predefined intervals. In the following Section, we conduct a thorough and systematic numerical exploration by revealing the Newton-Raphson basins of attraction and how they are affected by the value of the ratio $\lambda$. In Section \ref{conc}, the main conclusions of this work are presented, while our paper ends with Section \ref{futw} where we include a couple of interesting points for future work.

\section{Description of the mathematical model}
\label{mod}

\begin{figure}[!t]
\centering
\resizebox{\hsize}{!}{\includegraphics{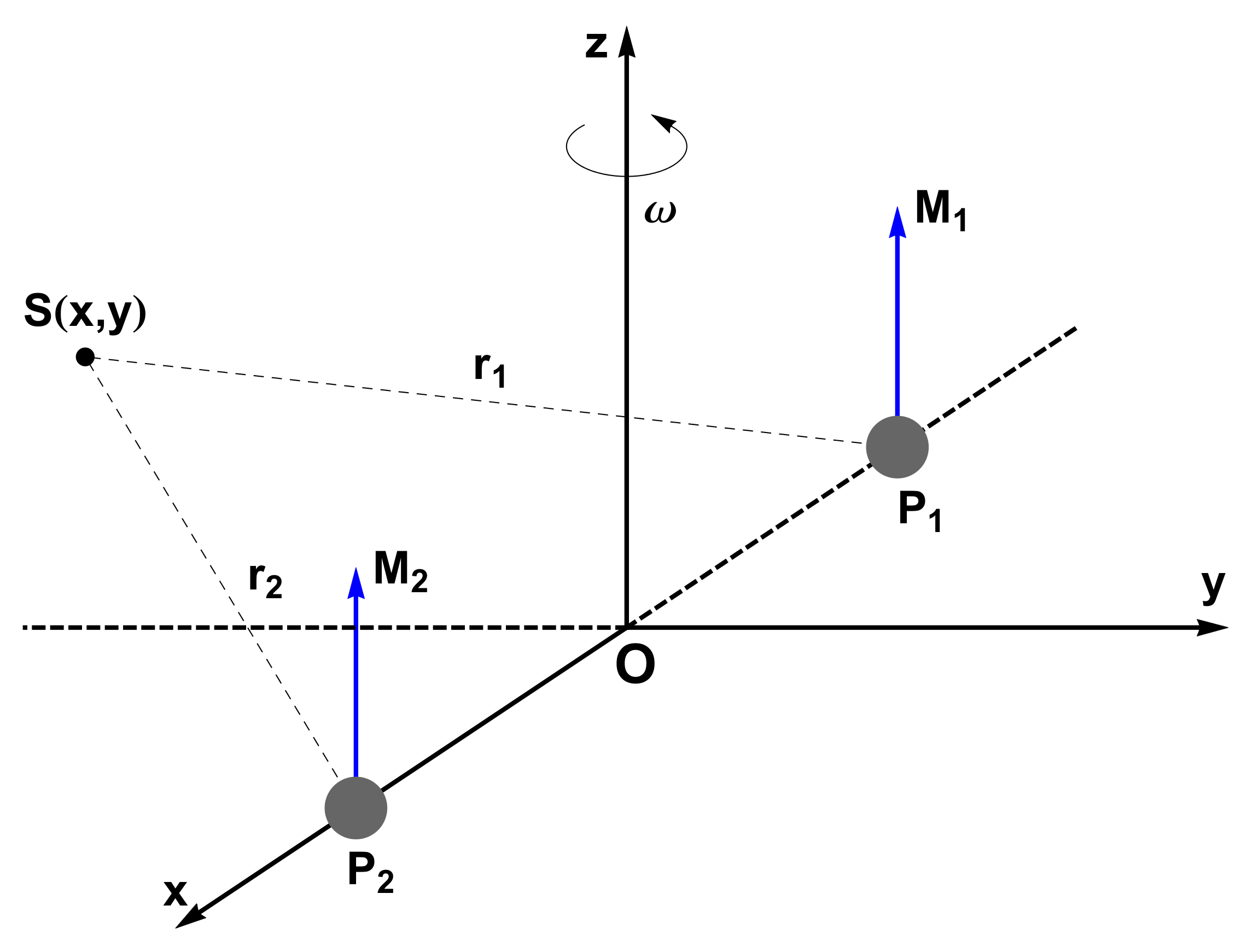}}
\caption{Configuration of the electromagnetic Copenhagen problem. The test particle $S$ moves on the $(x,y)$ plane}
\label{emgcp}
\end{figure}

Let us briefly recall the basic properties of the binary system with the magnetic dipoles. We assume that the two primaries rotate around their common center of mass in circular orbits, with constant angular velocity $\omega$, under their mutual Newtonian attraction. For the description of the system we use a synodic coordinates system $Oxyz$ where its center $O$ is located at the mass center of the system. Both primaries, which create dipole-type magnetic fields, move on the $Oxy$ plane, where the $Ox$ axis is the axis of syzygies of the primaries. It is assumed that the axes of the magnetic moments $M_i$ are both parallel to $Oz$ and perpendicular to the $Oxy$ plane, that is $\vec{M_i} = M_i(0,0,1)$, $i = 1, 2$ (see Fig. \ref{emgcp}).

In this synodic frame of reference the planar motion of a test particle can be described by the following set of differential equations
\begin{eqnarray}
U_x(x,y,z) &=& \frac{\partial U}{\partial x} = \ddot{x} - f \dot{y}, \nonumber\\
U_y(x,y,z) &=& \frac{\partial U}{\partial y} = \ddot{y} + f \dot{x}, \nonumber\\
U_z(x,y,z) &=& \frac{\partial U}{\partial z} = \ddot{z},
\label{eqmot}
\end{eqnarray}
where
\begin{equation}
U(x,y,z) = \omega\left(x A_y - y A_x\right) + \frac{\omega^2}{2}\left(x^2 + y^2\right),
\label{pot}
\end{equation}
is the potential function of the system, while $A_x$ and $A_y$ are the $x$ and $y$-components of the vector potential of the resultant electromagnetic field, expressed in the synodic frame of reference with
\begin{eqnarray}
A_x &=& - y p_1, \nonumber\\
A_y &=& x p_1 - q_1, \nonumber\\
p_1 &=& \frac{1}{r_1^3} + \frac{\lambda}{r_2^3}, \nonumber\\
q_1 &=& \frac{1}{2}\left(\frac{1}{r_1^3} - \frac{\lambda}{r_2^3}\right), \nonumber\\
p_2 &=& \frac{1}{r_1^5} + \frac{\lambda}{r_2^5}, \nonumber\\
q_2 &=& \frac{1}{2}\left(\frac{1}{r_1^5} - \frac{\lambda}{r_2^5}\right), \nonumber\\
s_2 &=& \frac{1}{4}\left(\frac{1}{r_1^5} + \frac{\lambda}{r_2^5}\right), \nonumber\\
f &=& 2\omega + 2p_1 - 3\left(x^2 + y^2\right)p_2 + 6xq_2 - 3s_2.
\label{eqs}
\end{eqnarray}

The distances $r_1$ and $r_2$ of the test particle from the two primaries are defined as
\begin{eqnarray}
r_1 &=& \sqrt{\left(x + \mu\right)^2 + y^2 + z^2}, \nonumber\\
r_2 &=& \sqrt{\left(x + \mu - 1\right)^2 + y^2 + z^2}.
\label{dist}
\end{eqnarray}

In this paper, we shall investigate the special case of the Copenhagen problem which means that the value of the mass ratio is $\mu = 1/2$. Moreover, the value of the angular velocity is $\omega = 1$.

Undoubtedly, the most outstanding parameter of the dynamical system is $\lambda = M_1/M_2$, which is of course the ratio of the magnitudes of the magnetic moments of the two primaries. This parameter directly connects the magnetic moments and therefore it plays a very important role on the formation of the resultant electromagnetic field.

The set of equations of motion (\ref{eqmot}) admits only the following Jacobian-type integral of motion
\begin{equation}
H\left(x,y,z,\dot{x},\dot{y},\dot{z}\right) = 2U(x,y,z) - \left(\dot{x}^2 + \dot{y}^2 + \dot{z}^2 \right) = C,
\end{equation}
where $C$ is its numerical value which remains constant. This constant is also know as the Jacobian constant or the energy constant.

\section{Parametric variation and stability of the equilibrium points}
\label{lgevol}

It is well known that in an equilibrium point the following conditions hold
\begin{equation}
\dot{x} = \dot{y} = \dot{z} = \ddot{x} = \ddot{y} = \ddot{z} = 0.
\label{lps1}
\end{equation}
Therefore, the coordinates of the positions of the equilibrium points can be numerically obtained by solving the system of non-linear algebraic differential equations
\begin{equation}
U_x = 0, \ U_y = 0, \ U_z = 0.
\label{lps}
\end{equation}

When the magnetic moments are perpendicular on the $(x,y)$ plane, where the two primaries revolve around their common center, then all the equilibrium points of the system lie on this particular plane (see e.g., \cite{KM87}). Some of the equilibrium points are dynamically equivalent and therefore they are characterized by the same state of stability and by the same value of the Jacobian constant $C$.

\begin{figure}[!t]
\centering
\resizebox{0.7\hsize}{!}{\includegraphics{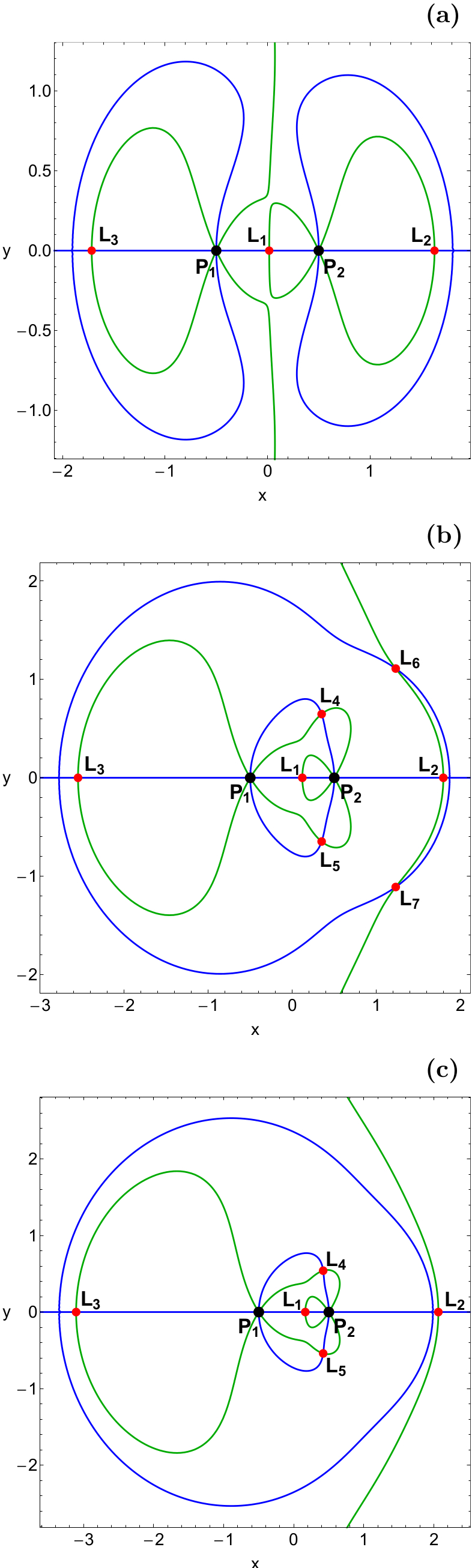}}
\caption{The red dots indicate the positions of the equilibrium points through the intersections of $U_x = 0$ (green) and $U_y = 0$ (blue), when (a-upper panel): $\lambda = 1.3$, (b-middle panel): $\lambda = 7$, and (c-lower panel): $\lambda = 15$. The black dots denote the centers of the two primaries.}
\label{lgs}
\end{figure}

The numerical value of the ratio $\lambda$ determines the total number of the existing equilibrium points. In particular, there are three cases regarding the total number of the equilibrium points. In Paper I the numerical ranges of $\lambda$ in each of the three cases were given. However the precision of the critical ending points was rather poor, so we decided to conduct some further numerical calculations in order to accurately determine the exact values of the ending points of the ranges. Our computations suggest that
\begin{itemize}
  \item When $0 < \lambda < \lambda_1$ there are only three collinear equilibrium points $(L_1, \ L_2, \ L_3)$ which all lie on the $x$-axis (see panel (a) of Fig. \ref{lgs}).
  \item When $\lambda_1 \leq \lambda \leq \lambda_2$ there are three collinear points $(L_1, \ L_2, \ L_3)$ and four triangular points that form two pairs $(L_4, \ L_5)$ and $(L_6, \ L_7)$. These equilibrium points are symmetric with respect to the $x$-axis and therefore they are dynamically equivalent (see panel (b) of Fig. \ref{lgs}).
  \item When $\lambda > \lambda_2$ there are three collinear points $(L_1, \ L_2, \ L_3)$ and only one pair of dynamically equivalent triangular points $(L_4, \ L_5)$ (see panel (c) of Fig. \ref{lgs}),
\end{itemize}
where $\lambda_1 = 2.57753104$ and $\lambda_2 = 10.88609773$\footnote{In Paper I both values of the critical ending points $\lambda_1$ and $\lambda_2$ were given with accuracy of only one decimal digit.}.

\begin{figure*}[!t]
\centering
\resizebox{\hsize}{!}{\includegraphics{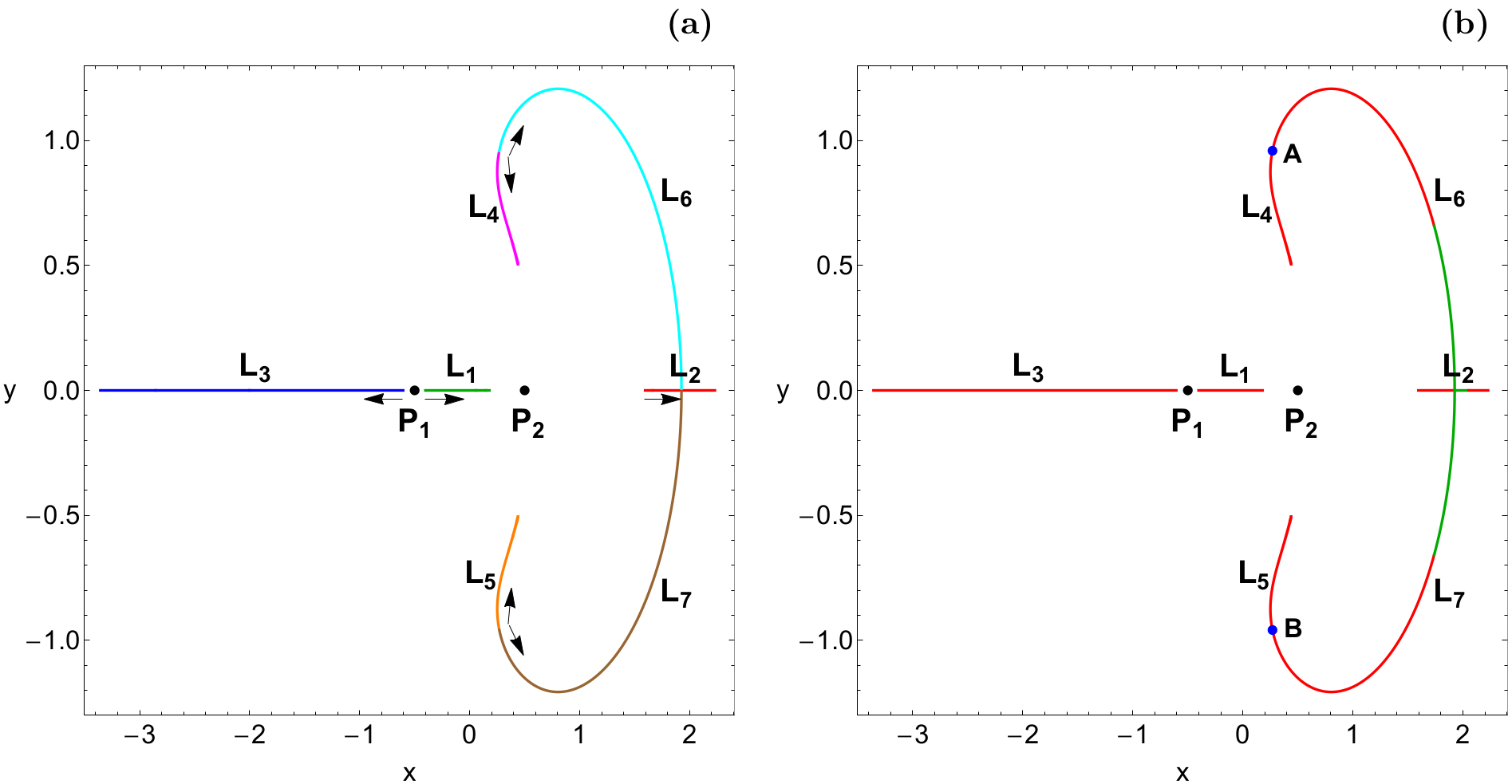}}
\caption{The space-evolution of (a-left): the position and (b-right): the stability (green) or instability (red) of the equilibrium points in the Copenhagen problem where the primaries are magnetic dipoles, when $\lambda \in (0, 20]$. The arrows indicate the movement direction of the equilibrium points as the ratio $\lambda$ of the magnetic moments increases. The black dots pinpoint the fixed centers of the two primaries, while the blue dots (points A and B) correspond to $\lambda_1$ where all four triangular points are born.}
\label{lgse}
\end{figure*}

In Fig. \ref{lgs} we see how the intersections of Eqs. $U_x = 0, \ U_y = 0$ define on the $(x,y)$ plane the positions of the equilibrium points when (a): $\lambda = 1.3$, (b): $\lambda = 7$, and (c): $\lambda = 15$.

In this paper, we shall explore how the ratio $\lambda$ of the magnetic moments influences the positions of the libration points, when it varies in the interval $\lambda \in (0, 20]$. Our results are illustrated in Fig. \ref{lgse}, where we present the space-evolution of the seven equilibrium points. One may observe that as value of the ratio $\lambda$ tends to zero $(\lambda \rightarrow 0)$ the libration points $L_1$ and $L_3$ move towards the center $P_1$. On the other hand, the equilibrium point $L_2$ moves away from the center $P_2$, with increasing value of $\lambda$. At the special case where $\lambda \rightarrow \lambda_1$ it was found that $L_4$ and $L_5$ tend to collide with $L_6$ and $L_7$, respectively. Furthermore, it it interesting to note that as the value of $\lambda$ tends to $\lambda_2$ both libration points $L_6$ and $L_7$ tend to collide with each other on the $x$-axis. Additional numerical calculations, for values of $\lambda$ beyond the interval $(0, 20]$, indicate that for $\lambda \rightarrow \infty$ the equilibrium points $L_1$, $L_4$, and $L_5$ tend asymptotically to center $P_2$. Before closing this section, we would like to point out that we did not consider cases with negative values of $\lambda$.

For determining the stability of an equilibrium point the origin of the frame of reference is transferred at its position $(x_0,y_0)$ following the transformation
\begin{align}
&x = x_0 + \xi, \nonumber\\
&y = y_0 + \eta.
\end{align}
Then we expand the system of equations of motion (\ref{eqmot}) into first-order terms with respect to $\xi$ and $\eta$ thus obtaining the linearized system which describes infinitesimal motions near an equilibrium point
\begin{equation}
\dot{{\bf{\Xi}}} = A {\bf{\Xi}}, \ \ {\bf{\Xi}} = \left(\xi, \eta, \dot{\xi}, \dot{\eta}\right)^{\rm T},
\label{ls}
\end{equation}
where ${\bf{\Xi}}$ is the state vector of the test particle with respect to the equilibrium points, while $A$ is the time-independent coefficient matrix of variations
\begin{equation}
A =
\begin{bmatrix}
    0 & 0 & 1 & 0 \\
    0 & 0 & 0 & 1 \\
    \Omega_{xx}^0 & \Omega_{xy}^0 & 0 & 2\omega \\
    \Omega_{yx}^0 & \Omega_{yy}^0 & -2\omega & 0
\end{bmatrix},
\end{equation}
where the superscript 0 at the partial derivatives of second order denotes evaluation at the position of the equilibrium point $(x_0, y_0)$.

The characteristic equation of the linear system (\ref{ls}) is quadratic with respect to $W = w^2$ and is given by
\begin{equation}
\alpha W^2 + b W + c = 0,
\label{ceq}
\end{equation}
where
\begin{align}
&\alpha = 1, \nonumber\\
&b = 4\omega^2 - \Omega_{xx}^0 - \Omega_{yy}^0, \nonumber\\
&c = \Omega_{xx}^0 \Omega_{yy}^0 - \Omega_{xy}^0 \Omega_{yx}^0.
\end{align}

An equilibrium point is stable only when all roots of the characteristic equation for $w$ are pure imaginary. Therefore the following three necessary and sufficient conditions must be simultaneously fulfilled
\begin{equation}
b > 0, \ \ c > 0, \ \ D = b^2 - 4 a c > 0,
\end{equation}
which ensure that the characteristic equation (\ref{ceq}) has two real negative roots $W_{1,2}$, which consequently means that there are four pure imaginary roots for $w$.

Knowing the exact position of the equilibrium points (see panel (a) of Fig. \ref{lgse}) we can easily insert them into Eq. (\ref{ceq}), determine the nature of the fours roots and then derive the stability of the libration points. Our numerical calculations suggest that the collinear points (for which $\Omega_{xy}^0 = \Omega_{yx}^0 = 0$) $L_1$ and $L_3$, as well as the triangular points $L_4$ and $L_5$ are unstable for all values of the ratio of the magnitudes of the magnetic moments in the interval $\lambda \in (0,20]$. The equilibrium points $L_2$, $L_6$, and $L_7$ on the other hand, can be either stable or unstable, depending of course on the particular value of $\lambda$. In panel (b) of Fig. \ref{lgse} we present the evolution of the stability of all the equilibrium points, when $\lambda \in (0,20]$. Our numerical calculations suggest that the collinear point $L_2$ is stable when $\lambda \in [\lambda_2,14.702571987]$, while the triangular points $L_6$ and $L_7$ are stable only when $\lambda$ lines in the interval $(9.79758826,\lambda_2]$.

\section{The Newton-Raphson basins of attraction}
\label{bas}

In Paper I the Newton-Raphson scheme as well as the modified Broyden's method were used for determining the corresponding basins of convergence. It was shown that the Newton-Raphson iterative scheme is much faster than the modified Broyden's scheme. Moreover, in the modified Broyden's method a substantial number of non-converging points exist, while on the other hand in the Newton-Raphson scheme the same type of initial conditions are very limited. Therefore, we decided to choose the Newton-Raphson method and conduct a thorough and systematic investigation on the basins of attraction in the electromagnetic binary system.

The Newton-Raphson method is applicable to systems of multivariate functions $f({\bf{x}}) = 0$, through the iterative scheme \begin{equation}
{\bf{x}}_{n+1} = {\bf{x}}_{n} - J^{-1}f({\bf{x}}_{n}),
\label{sch}
\end{equation}
where $J^{-1}$ is the inverse Jacobian matrix of $f({\bf{x_n}})$. In our case the system of differential equations is
\begin{equation}
\begin{cases}
U_x = 0 \\
U_y = 0
\end{cases},
\label{sys}
\end{equation}
and therefore the Jacobian matrix reads
\begin{equation}
J =
\begin{bmatrix}
U_{xx} & U_{xy} \\
U_{yx} & U_{yy}
\end{bmatrix}.
\label{jac}
\end{equation}
The inverse Jacobian is
\begin{equation}
J^{-1} = \frac{1}{{\rm{det}}(J)}
\begin{bmatrix}
U_{yy} & - U_{xy} \\
- U_{yx} & U_{xx}
\end{bmatrix},
\label{ijac}
\end{equation}
where ${\rm{det}}(J) = U_{yy} U_{xx} - U_{xy}^2$.

Inserting the expression of the inverse Jacobian into the iterative formula (\ref{sch}) we get
\begin{eqnarray}
\begin{bmatrix}
x \\
y
\end{bmatrix}
_{n+1} &=&
\begin{bmatrix}
x \\
y
\end{bmatrix}
_{n} - \frac{1}{U_{yy} U_{xx} - U_{xy}^2}
\begin{bmatrix}
U_{yy} & - U_{xy} \\
- U_{yx} & U_{xx}
\end{bmatrix}
\begin{bmatrix}
U_x \\
U_y
\end{bmatrix}
_{(x_n,y_n)}
\nonumber\\
&=&
\begin{bmatrix}
x \\
y
\end{bmatrix}
_{n} - \frac{1}{U_{yy} U_{xx} - U_{xy}^2}
\begin{bmatrix}
U_{yy} U_x - U_{xy} U_y \\
- U_{yx} U_x + U_{xx} U_y
\end{bmatrix}
_{(x_n,y_n)}.
\label{sch2}
\end{eqnarray}

Decomposing formula (\ref{sch2}) into $x$ and $y$ we obtain the iterative formulae for each coordinate
\begin{eqnarray}
x_{n+1} &=& x_n - \left( \frac{U_x U_{yy} - U_y U_{xy}}{U_{yy} U_{xx} - U^2_{xy}} \right)_{(x_n,y_n)}, \nonumber\\
y_{n+1} &=& y_n + \left( \frac{U_x U_{yx} - U_y U_{xx}}{U_{yy} U_{xx} - U^2_{xy}} \right)_{(x_n,y_n)},
\label{nrm}
\end{eqnarray}
where $x_n$, $y_n$ are the values of the $x$ and $y$ variables at the $n$-th step of the iterative process, while the subscripts of $U$ denote the corresponding partial derivatives of the potential function $U(x,y,z = 0)$.

The Newton-Raphson algorithm is activated when an initial condition $(x_0,y_0)$ on the configuration plane is given, while it stops when the positions of the equilibrium points are reached, with some predefined accuracy. A crooked path line is usually created by the successive approximations-points. If the iterative method converges for the particular initial conditions then this path leads to a desired position, which in our case is one of the equilibrium points. All the initial conditions that lead to a specific equilibrium point, compose a basin of attraction/convergence\footnote{It should be clarified that the Newton-Raphson basins of attraction should not be mistaken with the classical basins of attraction in dissipative systems.} or an attracting region/domain. We observe that the iterative formulae (\ref{nrm}) include both the first and the second derivatives of the effective potential function $U(x,y,z = 0)$ and therefore we may claim that the obtained numerical results directly reflect some of the basic qualitative characteristics of the dynamical system. The major advantage of knowing the Newton-Raphson basins of attraction in a dynamical system is the fact that we can select the most favorable initial conditions, with respect to required computation time, when searching for an equilibrium point.

For obtaining the basins of convergence we worked as follows: First we defined a dense uniform grid of $1024 \times 1024$ initial conditions (nodes) regularly distributed on the configuration $(x,y)$ space which will be used as initial values of the numerical algorithm. The iterative process was terminated when an accuracy of $10^{-15}$ has been reached, which is almost double of the accuracy used in Paper I. We classified all the $(x_0,y_0)$ initial conditions that lead to a particular solution (equilibrium point) by performing a double scanning process of the configuration $(x,y)$ plane. At the same time, for each initial point, we recorded the number $(N)$ of iterations required to obtain the aforementioned accuracy. Logically, the required number of iterations for locating an equilibrium point strongly depends on the value of the predefined accuracy. In this study we set the maximum number of iterations $N_{\rm max}$ to be qual to 500.

All the computations reported in this paper regarding the basins of attraction were performed using a double precision algorithm written in standard \verb!FORTRAN 77! (e.g., \cite{PTVF92}). Furthermore, all graphical illustrations have been created using the latest version 11 of Mathematica$^{\circledR}$ \cite{W03}.

In the following we shall try to determine how the ratio $\lambda$ of the magnetic moments influences the Newton-Raphson basins of attraction, considering three cases regarding the total number of the equilibrium points.

\subsection{Case I: $0 < \lambda < \lambda_1$ (three equilibrium points)}
\label{ss1}

\begin{figure*}[!t]
\centering
\resizebox{0.7\hsize}{!}{\includegraphics{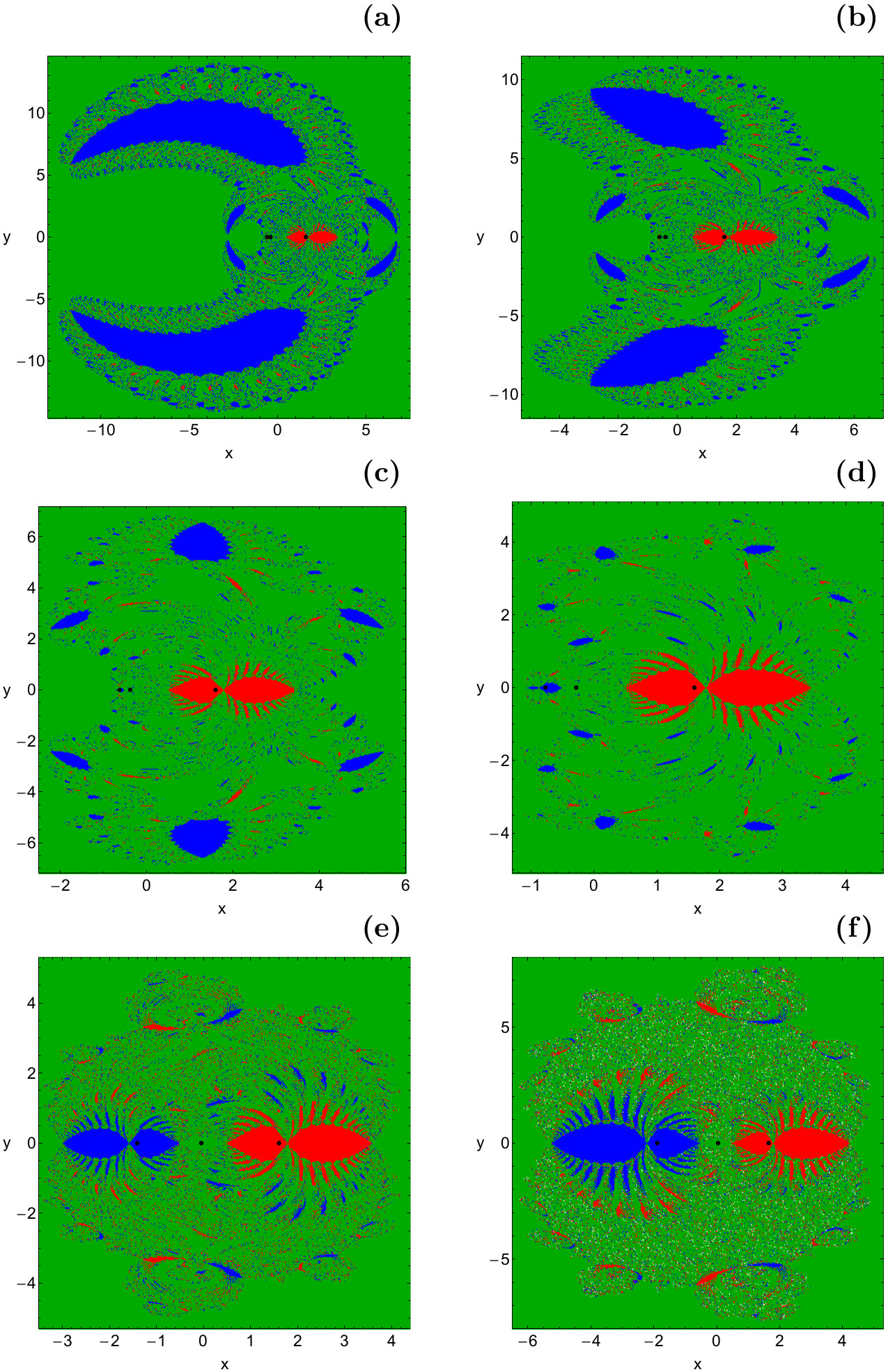}}
\caption{The Newton-Raphson basins of attraction on the configuration $(x,y)$ plane for the first case, where three equilibrium points are present. (a): $\lambda = 0.0005$; (b): $\lambda = 0.00055$; (c): $\lambda = 0.001$; (d): $\lambda = 0.01$; (e): $\lambda = 0.5$; (f): $\lambda = 2.57$. The positions of the three equilibrium points are indicated by black dots. The color code denoting the three attractors is as follows: $L_1$ (green); $L_2$ (red); $L_3$ (blue). The non-converging points are indicated with white color.}
\label{3lgp}
\end{figure*}

\begin{figure*}[!t]
\centering
\resizebox{0.7\hsize}{!}{\includegraphics{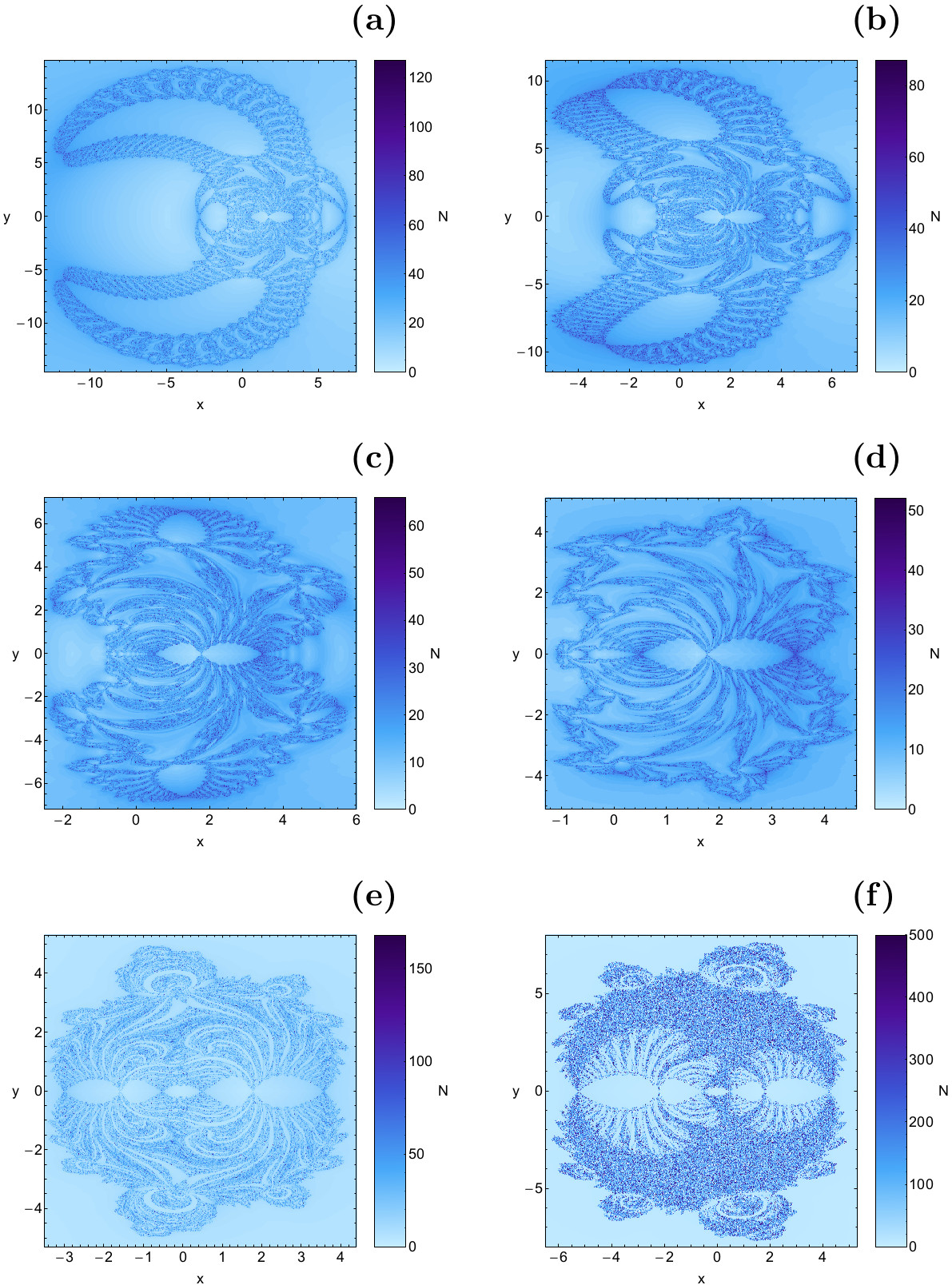}}
\caption{The distribution of the corresponding number $(N)$ of required iterations for obtaining the Newton-Raphson basins of attraction shown in Fig. \ref{3lgp}(a-f).}
\label{3lgpn}
\end{figure*}

\begin{figure*}[!t]
\centering
\resizebox{0.7\hsize}{!}{\includegraphics{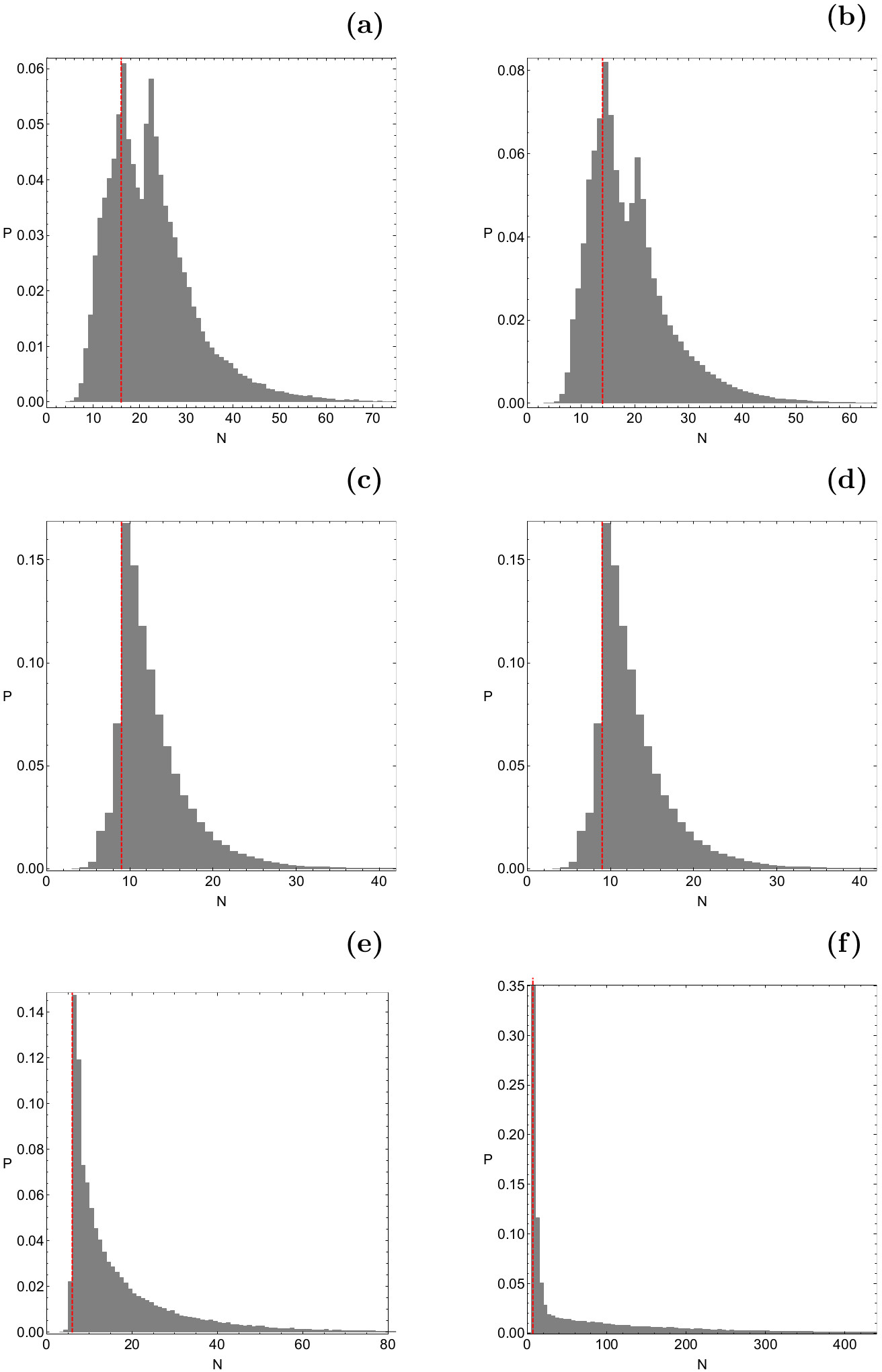}}
\caption{The corresponding probability distribution of required iterations for obtaining the Newton-Raphson basins of attraction shown in Fig. \ref{3lgp}(a-f). The vertical dashed red line indicates, in each case, the most probable number $(N^{*})$ of iterations.}
\label{3lgpp}
\end{figure*}

Our exploration begins with the case where three collinear equilibrium points are present. In Fig. \ref{3lgp}(a-f) we present the Newton-Raphson basins of attraction for six values of the ratio $\lambda$ of the magnetic moments. It is interesting to note that the basins of convergence corresponding to the central point $L_1$ extend to infinity, while on the other hand the area of the other two basins of attraction is finite. It is evident that a large portion of the configuration $(x,y)$ plane is covered by well-formed basins of attraction. The boundaries between the several basins of convergence however are, in many cases, highly fractal\footnote{When we state that an area is fractal we simply mean that it has a fractal-like geometry without conducting, at least for now, any specific calculations for computing the fractal dimensions as in \cite{AVS09}.} and they look like a ``chaotic sea". This means that if we choose a starting point $(x_0,y_0)$ of the Newton-Raphson method inside these fractal domains we will observe that our choice is very sensitive. More precisely, a slight change in the initial conditions leads to completely different final destination (different attractor) and therefore the beforehand prediction becomes extremely difficult.

Looking the color-coded plots we may say that the shape of the basin of attraction corresponding to equilibrium point $L_2$ has the shape of an exotic bug with many legs and many antennas. The same also applies for the basin of attraction of libration point $L_3$ however only for relatively high values of $\lambda$.

In Fig. \ref{3lgpn}(a-f) we provide the distribution of the corresponding number $(N)$ of iterations required for obtaining the desired accuracy, using tones of blue. In the same vein, in Fig. \ref{3lgpp}(a-f) the corresponding probability distribution of iterations is shown. The probability $P$ is defined as follows: let us assume that $N_0$ initial conditions $(x_0,y_0)$ converge to one of the five attractors after $N$ iterations. Then $P = N_0/N_t$, where $N_t$ is the total number of initial conditions in every grid. Table \ref{table1} contains some interesting results regarding the percentages of the basins of attraction, the required number of iterations and the non-converging points. It should be noted that the percentage of the basins of attraction corresponding to $L_1$ is not included because these basins extend to infinity and therefore the percentage has no meaning (it depends on the particular size of the rectangular grid).

\begin{table}
\begin{center}
   \caption{The percentages of the Newton-Raphson basins of attraction and the non-converging points (NC) along with the most probable number $N^{*}$ of iterations, for the case where three equilibrium points exist.}
   \label{table1}
   \setlength{\tabcolsep}{10pt}
   \begin{tabular}{@{}ccccl}
      \hline
      $\lambda$ & $L_2$ (\%) & $L_3$ (\%) & NC (\%) & $N^{*}$ \\
      \hline
      0.00050 & 1.86 & 20.66 & 0.00 & 16 \\
      0.00055 & 2.56 & 15.01 & 0.00 & 14 \\
      0.00100 & 3.68 &  6.51 & 0.00 & 11 \\
      0.01000 & 5.68 &  2.98 & 0.00 &  9 \\
      0.05000 & 6.48 &  4.47 & 0.00 &  8 \\
      0.50000 & 7.80 &  6.01 & 0.00 &  7 \\
      1.00000 & 7.85 &  6.18 & 0.33 &  6 \\
      1.50000 & 7.87 &  6.98 & 0.00 &  6 \\
      2.00000 & 8.43 &  7.49 & 0.00 &  6 \\
      2.57000 & 8.09 &  7.67 & 1.42 &  6 \\
      \hline
   \end{tabular}
\end{center}
\end{table}

Correlating all the numerical results given in Figs. \ref{3lgp}, \ref{3lgpn}, and \ref{3lgpp}, as well as in Table \ref{table1} one may conclude that the most important phenomena which take place as the value of $\lambda$ increases are the following:
\begin{itemize}
  \item Initially, when the value of the ratio $\lambda$ is just above zero, the shape of the basins of attraction corresponding to libration point $L_3$ have a wing-type shape. However, as we proceed to higher values of $\lambda$ their shape starts to change and finally it resembles the bug shape of basins of attraction of equilibrium point $L_2$.
  \item The area of basins of convergence corresponding to $L_2$ is constantly increasing, while on the other hand the evolution of the area of basins of attraction of $L_3$ is not monotonic. In particular, it decreases until $\lambda = 0.01$, while for higher values of the ratio of the magnetic moments the tendency is reversed.
  \item The average value of required number $(N)$ of iterations for obtaining the desired accuracy decreases. Consequently, the the most probable number $(N^{*})$ of iterations is reduced from 16 when $\lambda = 0.0005$ to 6 when $\lambda = 2.57$.
  \item For $\lambda > 0.5$ it was found that the range of the probability distribution of the required iterations significantly increases. Indeed, we observe in panel (f) of Fig. \ref{3lgpp} that for $\lambda = 2.57$ almost the entire range of available iterations, $N \in [0, 500]$, is occupied.
  \item A small non-zero amount of non-converging points was found in the case where the magnitudes of the magnetic moments are equal $(\lambda = 1)$. Furthermore, our computations indicate that as we approach the limiting ending point $\lambda_1$ the rate of the non-converging points increases.
\end{itemize}

\subsection{Case II: $\lambda_1 \leq \lambda \leq \lambda_2$ (seven equilibrium points)}
\label{ss2}

\begin{figure*}[!t]
\centering
\resizebox{0.9\hsize}{!}{\includegraphics{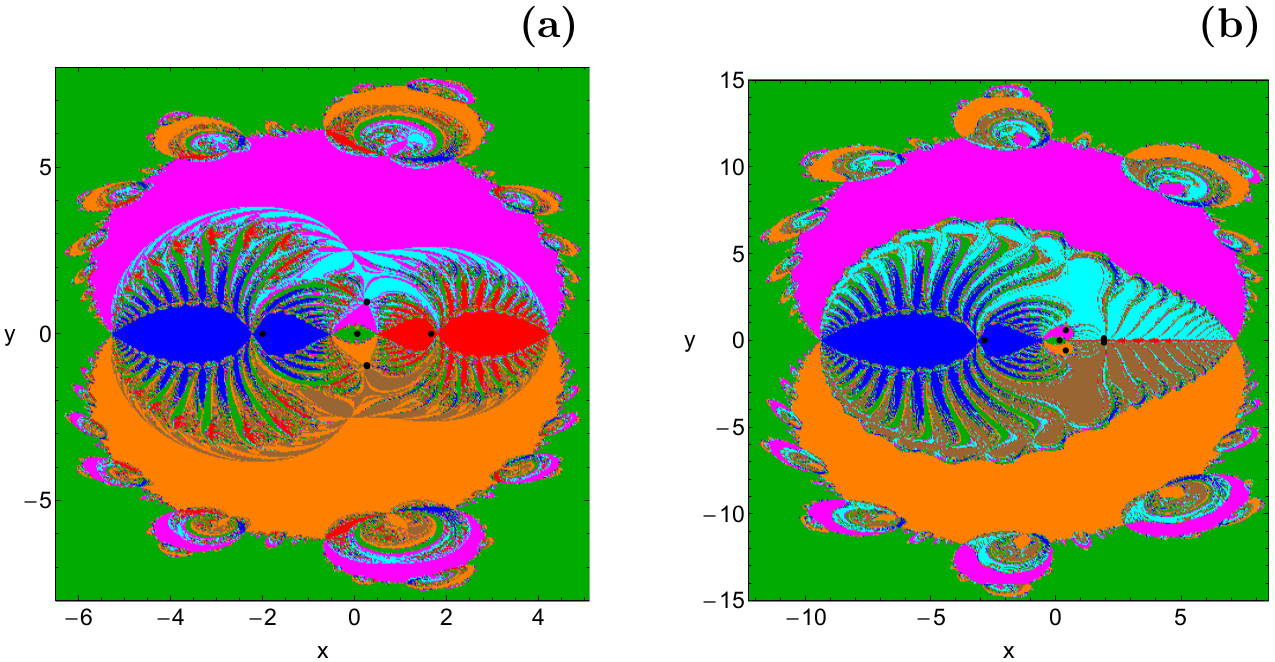}}
\caption{The Newton-Raphson basins of attraction on the configuration $(x,y)$ plane for the second case, where seven equilibrium points are present. (a-left): $\lambda = 2.58$; (b-right): $\lambda = 10.85$. The positions of the seven equilibrium points are indicated by black dots. The color code denoting the seven attractors is as follows: $L_1$ (green); $L_2$ (red); $L_3$ (blue); $L_4$ (magenta); $L_5$ (orange); $L_6$ (cyan); $L_7$ (brown). The non-converging points are indicated with white color.}
\label{7lgp}
\end{figure*}

\begin{figure*}[!t]
\centering
\resizebox{0.9\hsize}{!}{\includegraphics{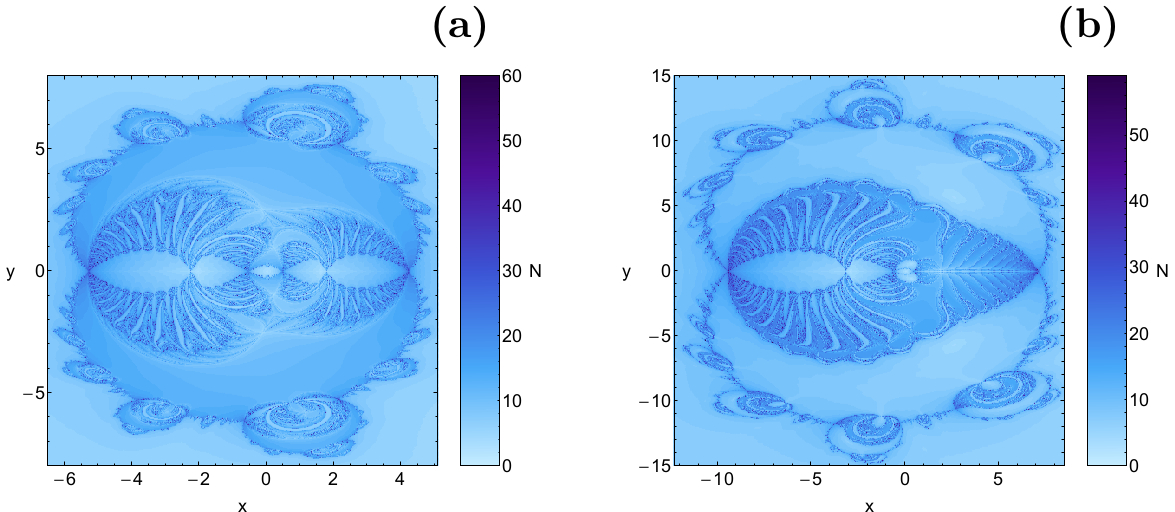}}
\caption{The distribution of the corresponding number $(N)$ of required iterations for obtaining the Newton-Raphson basins of attraction shown in Fig. \ref{7lgp}(a-b).}
\label{7lgpn}
\end{figure*}

\begin{figure*}[!t]
\centering
\resizebox{0.9\hsize}{!}{\includegraphics{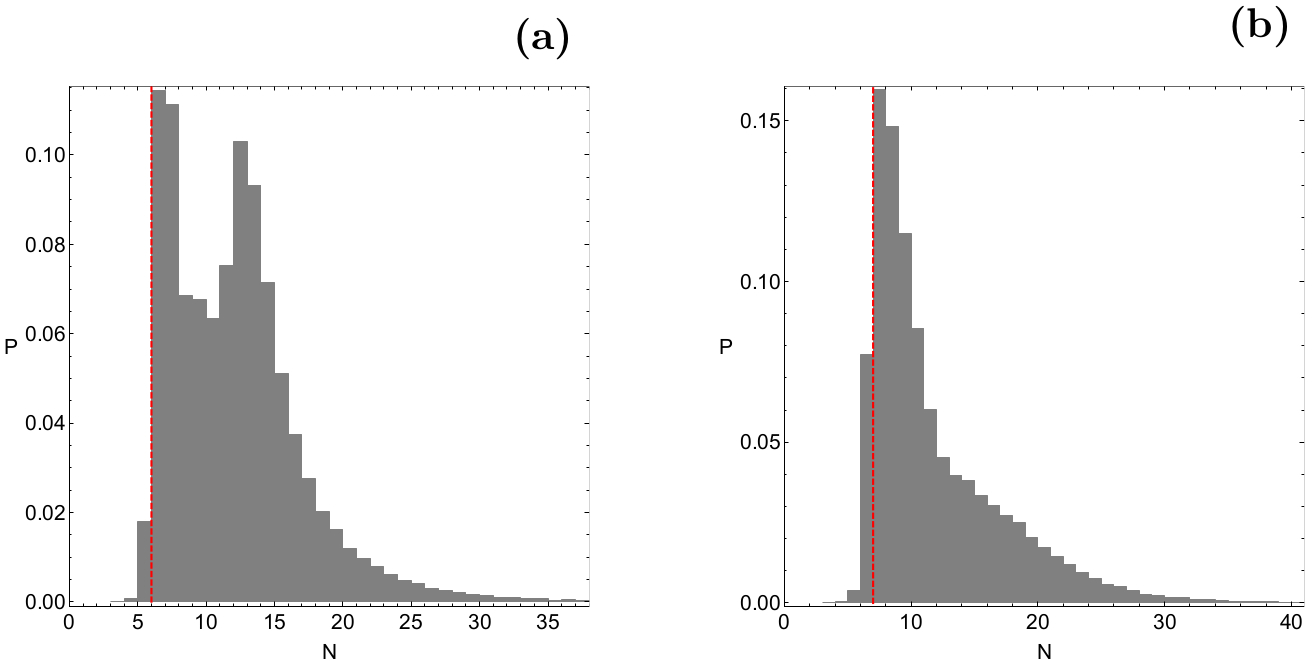}}
\caption{The corresponding probability distribution of required iterations for obtaining the Newton-Raphson basins of attraction shown in Fig. \ref{7lgp}(a-b). The vertical dashed red line indicates, in each case, the most probable number $(N^{*})$ of iterations.}
\label{7lgpp}
\end{figure*}

We continue our investigation with the case where seven libration points are present. The Newton-Raphson basins of attraction for two values of the ratio $\lambda$ are presented in Fig. \ref{7lgp}(a-b). In this range of values of $\lambda$ ($\lambda_1 \leq \lambda \leq \lambda_2$) the changes on the configuration $(x,y)$ plane due to the variation of the value of the ratio of the magnetic moments are not so prominent as in the previous case. Thus we decided to present in Fig. \ref{7lgp} only two (instead of six) color-coded convergence diagrams. Moreover, Fig. \ref{7lgpn}(a-b) shows the distribution of the corresponding number $(N)$ of iterations required for obtaining the desired accuracy. The corresponding probability distribution of iterations is given in Fig. \ref{7lgpp}(a-b). In Table \ref{table2} we provide some interesting results regarding the rates of the basins of convergence, the required number of iterations as well as the non-converging points.

One may observe that the shape of the main basins of attraction corresponding to equilibrium points $L_4$ and $L_5$ look like butterfly wings and they surround the attracting domains of the equilibrium points $L_2$ and $L_3$. On the other hand, the regions of convergence of libration points $L_6$ and $L_7$ are smaller and they are mainly confined between the basins of attraction of $L_4$ and $L_5$.

\begin{table}
\begin{center}
   \caption{The percentages of the Newton-Raphson basins of attraction and the non-converging points (NC) along with the most probable number $N^{*}$ of iterations, for the case where seven equilibrium points exist.}
   \label{table2}
   \setlength{\tabcolsep}{7.0pt}
   \begin{tabular}{@{}ccccccl}
      \hline
      $\lambda$ & $L_2$ (\%) & $L_3$ (\%) & $L_4 (\%)$ & $L_6 (\%)$ & NC (\%) & $N^{*}$ \\
      \hline
       2.58 & 5.26 & 5.74 & 18.87 & 5.48 & 0.00 & 6 \\
       4.24 & 5.07 & 6.08 & 19.89 & 6.91 & 0.00 & 7 \\
       5.88 & 4.59 & 6.37 & 19.17 & 6.98 & 0.00 & 7 \\
       7.52 & 4.19 & 6.60 & 18.87 & 7.30 & 0.00 & 7 \\
       9.16 & 3.21 & 6.65 & 17.52 & 7.44 & 0.00 & 7 \\
      10.85 & 0.77 & 6.74 & 18.25 & 8.95 & 0.00 & 8 \\
      \hline
   \end{tabular}
\end{center}
\end{table}

Taking into account all the numerical outcomes given in Figs. \ref{7lgp}, \ref{7lgpn}, and \ref{7lgpp}, as well as in Table \ref{table2} one may reasonably deduce that the most significant phenomena which take place as the value of $\lambda$ increases are the following:
\begin{itemize}
  \item The area of the basins of convergence corresponding to equilibrium point $L_3$ slightly increases, while that of $L_4$ exhibits small fluctuations around 18\%.
  \item The area of attracting domains of libration points $L_2$, $L_6$, and $L_7$ are much more influenced by the change on the value of the ratio $\lambda$. It is seen that as we proceed to higher values of $\lambda$ the basins of attraction of both $L_6$ and $L_7$ strongly suppress the bug-like attracting region of $L_2$ which finally completely disappears.
  \item The average value of required number $(N)$ of iterations for obtaining the desired accuracy remains almost unperturbed. More precisely, the the most probable number $(N^{*})$ of iteration starts at 6 for $\lambda = 2.58$, then it remains constant at 7, while for $\lambda = 10.85$ it was found slightly elevated at 8.
  \item In all examined cases, for more than 95\% of the initial conditions on the configuration $(x,y)$ plane the iterative formulae (\ref{nrm}) need no more than 35 iterations for obtaining the desired accuracy.
  \item In this range of values of the ratio of the magnetic moments there was absolutely no numerical indication of non-converging points.
\end{itemize}

\subsection{Case III: $\lambda > \lambda_2$ (five equilibrium points)}
\label{ss3}

\begin{figure*}[!t]
\centering
\resizebox{0.9\hsize}{!}{\includegraphics{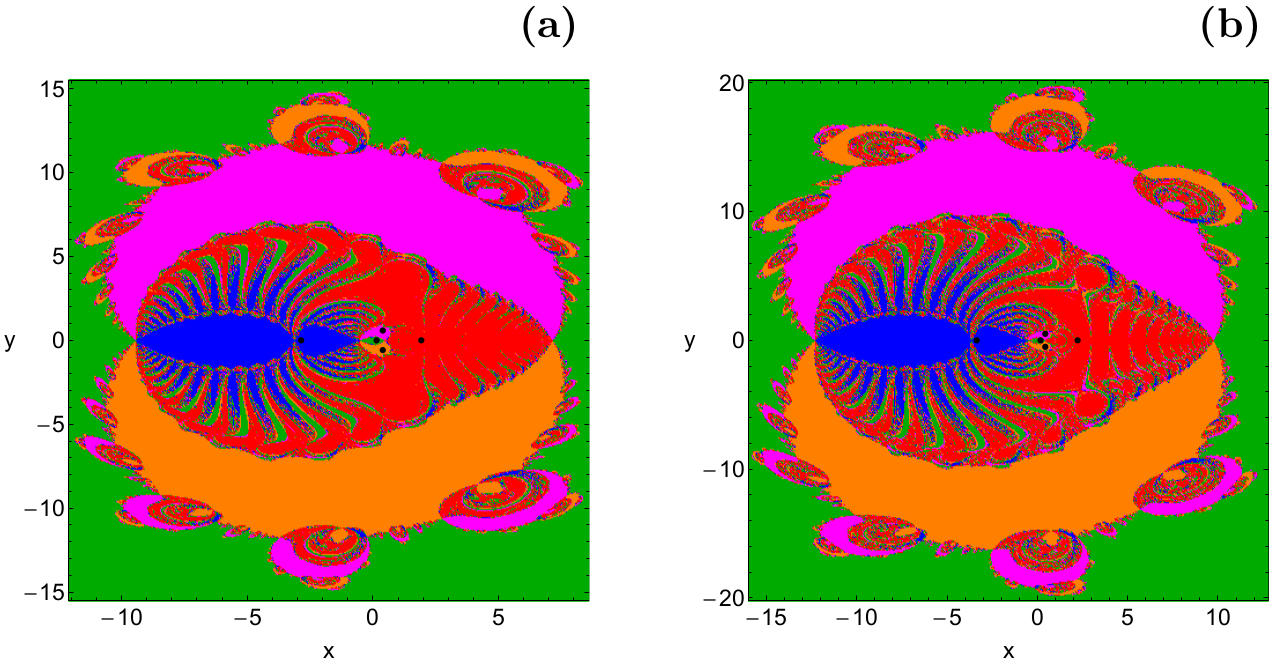}}
\caption{The Newton-Raphson basins of attraction on the configuration $(x,y)$ plane for the third case, where five equilibrium points are present. (a-left): $\lambda = 10.89$; (b-right): $\lambda = 20.00$. The positions of the five equilibrium points are indicated by black dots. The color code denoting the five attractors is as follows: $L_1$ (green); $L_2$ (red); $L_3$ (blue); $L_4$ (magenta); $L_5$ (orange). The non-converging points are indicated with white color.}
\label{5lgp}
\end{figure*}

\begin{figure*}[!t]
\centering
\resizebox{0.9\hsize}{!}{\includegraphics{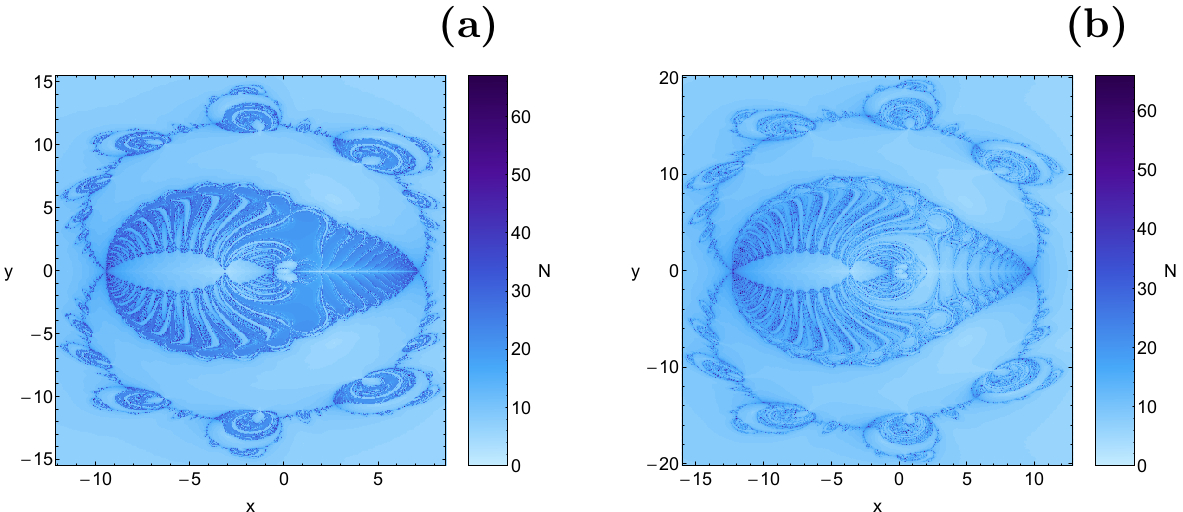}}
\caption{The distribution of the corresponding number $(N)$ of required iterations for obtaining the Newton-Raphson basins of attraction shown in Fig. \ref{5lgp}(a-b).}
\label{5lgpn}
\end{figure*}

\begin{figure*}[!t]
\centering
\resizebox{0.9\hsize}{!}{\includegraphics{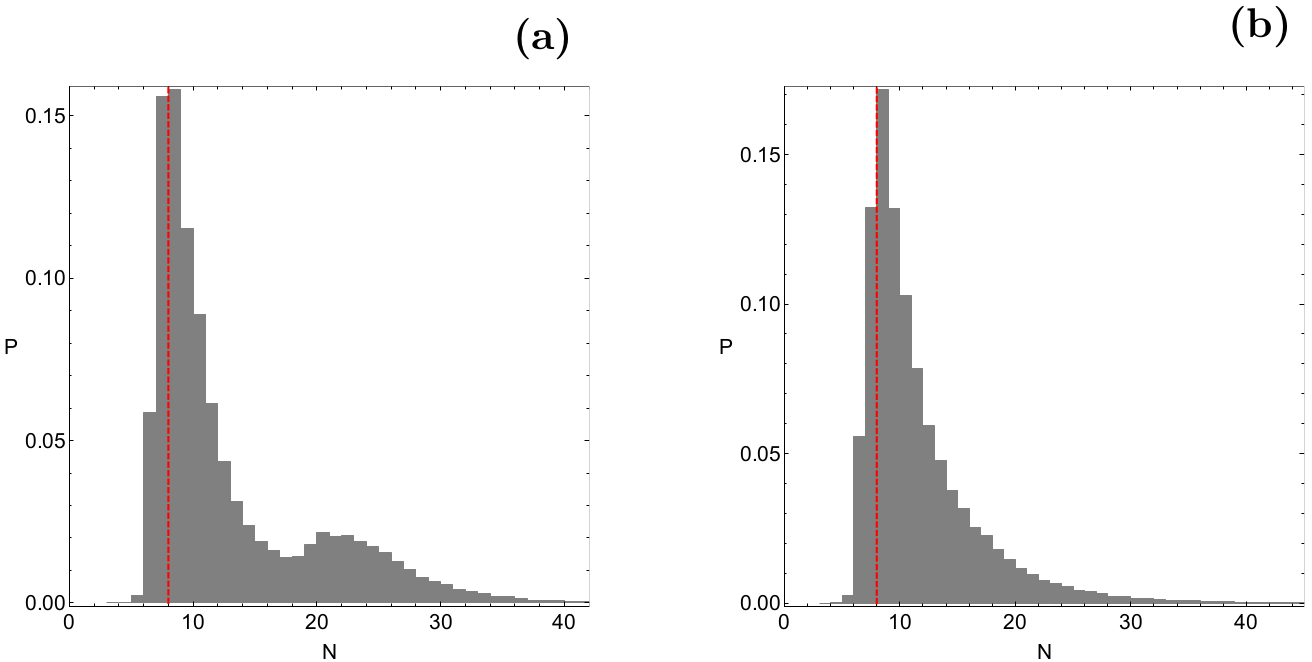}}
\caption{The corresponding probability distribution of required iterations for obtaining the Newton-Raphson basins of attraction shown in Fig. \ref{5lgp}(a-b). The vertical dashed red line indicates, in each case, the most probable number $(N^{*})$ of iterations.}
\label{5lgpp}
\end{figure*}

The last case under consideration concerns the scenario according to which the system of the two magnetic dipoles contains five equilibrium points. In Fig. \ref{5lgp}(a-b) we illustrate the Newton-Raphson basins of convergence for two values of the ratio $\lambda$. The corresponding number $(N)$ of required iterations for the desired accuracy is shown in Fig. \ref{5lgpn}(a-b), while the probability distribution of iterations is presented in Fig. \ref{5lgpp}(a-b). Some useful information regarding the percentages of the basins of attraction, the required number of iterations as well as the non-converging points is provided in Table \ref{table3}.

\begin{table}
\begin{center}
   \caption{The percentages of the Newton-Raphson basins of attraction and the non-converging points (NC) along with the most probable number $N^{*}$ of iterations, for the case where five equilibrium points exist.}
   \label{table3}
   \setlength{\tabcolsep}{10pt}
   \begin{tabular}{@{}cccccl}
      \hline
      $\lambda$ & $L_2$ (\%) & $L_3$ (\%) & $L_4 (\%)$ & NC (\%) & $N^{*}$ \\
      \hline
      10.89 & 18.17 & 6.60 & 17.73 & 0.00 & 8 \\
      12.00 & 18.17 & 6.73 & 17.79 & 0.00 & 8 \\
      14.00 & 17.35 & 6.59 & 17.32 & 0.00 & 8 \\
      16.00 & 17.25 & 6.78 & 17.45 & 0.00 & 8 \\
      18.00 & 16.88 & 7.00 & 17.85 & 0.00 & 8 \\
      20.00 & 16.01 & 7.09 & 18.06 & 0.00 & 8 \\
      \hline
   \end{tabular}
\end{center}
\end{table}

Summarizing all the results presented in Figs. \ref{5lgp}, \ref{5lgpn}, and \ref{5lgpp}, as well as in Table \ref{table3} we may conclude that the most considerable phenomena which take place as the value of $\lambda$ increases are the following:
\begin{itemize}
  \item In this case, we could argue that the structure of the basins of convergence does not practically change. What really happens is that as the value of $\lambda$ increases the entire pattern on the configuration $(x,y)$ plane grows in size. However the relative percentages of the attracting regions indicate that the $L_2$ and the $L_3$ domains exhibit a slight decrease and increase, respectively.
  \item It is interesting to note that in this range of values of the ratio of the magnetic moments the shape of the main attracting domain corresponding to libration point $L_2$ is different, with respect to the two previous cases. Indeed, the bug-like shape does not exist any more.
  \item The average value of required number $(N)$ of iterations for obtaining the desired accuracy remains constant. This means that the most probable number of iteration is $N^{*} = 8$ throughout the range of values of $\lambda$.
  \item In all studied cases, for more than 95\% of the examined initial conditions on the configuration $(x,y)$ plane the iterative scheme (\ref{nrm}) needs no more than 40 iterations for obtaining the desired accuracy.
  \item Our numerical calculations suggest that for $\lambda > \lambda_2$ there are no initial conditions on the $(x,y)$ plane which do not converge, sooner or later, to any of the five equilibrium points (attractors).
\end{itemize}

\subsection{An overview analysis}
\label{geno}

\begin{figure*}[!t]
\centering
\resizebox{0.9\hsize}{!}{\includegraphics{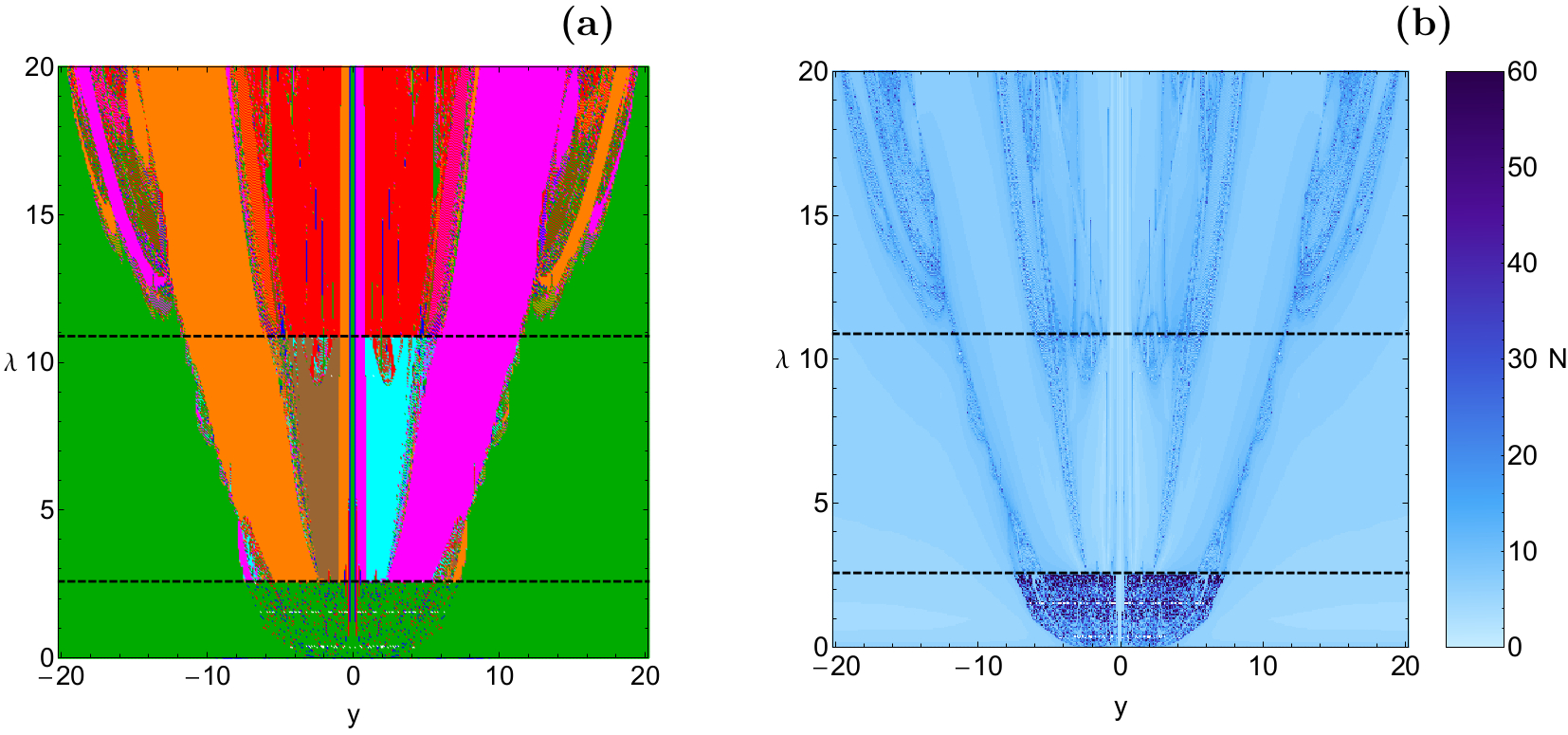}}
\caption{(a-left): The Newton-Raphson basins of attraction on the $(y,\lambda)$ plane, when $\lambda \in (0,20]$. The color code denoting the seven attractors is the same as in Fig. \ref{7lgp}. The horizontal black dashed lines indicate the critical ending points $\lambda_1$ and $\lambda_2$. (b-right): The distribution of the corresponding number $(N)$ of required iterations for obtaining the Newton-Raphson basins of attraction shown in panel (a).}
\label{yl}
\end{figure*}

The color-coded convergence diagrams on the configuration $(x,y)$ space provide sufficient information regarding the attracting domains however for only a fixed value of the ratio $\lambda$. In order to overcome this problem we can define another type of initial conditions which will allow us to scan a continuous spectrum of $\lambda$ values rather than few discrete levels. The most easy configuration is to set on the two coordinates $(x,y)$ equal to zero. In this case we choose to set $x = 0$, while the initial value of $y$ will vary in the interval $[-20,20]$. We could also choose to set $y = 0$ and let $x$ coordinate vary in the same interval. However, additional numerical calculations strongly suggest that the structure of the $(y,\lambda)$ plane is much more interesting than that of the $(x,\lambda)$ plane. This technique allows us to construct, once more, a two-dimensional plane in which the $y$ coordinate of orbits is the abscissa, while the value of $\lambda$ is the ordinate. In panel (a) of Fig. \ref{yl} we present the attracting domains of the $(y,\lambda)$ plane when $\lambda \in (0,20]$, while in panel (b) of the same figure the distribution of the corresponding number $(N)$ of required iterations for obtaining the Newton-Raphson basins of attraction is shown. The critical ending points $\lambda_1$ and $\lambda_2$ are indicated using horizontal black dashed lines.

We observe that in the first interval regarding the values of $\lambda$, that is $0 < \lambda < \lambda_1$, there are no well-formed basins of attraction. What exists is a highly fractal mixture of initial conditions converging to either of the three attractors. In the same interval we detected, for several values of the ratio of the magnetic moments, a tiny amount (about 0.05\%) of non-converging points. For $\lambda \geq \lambda_1$ the structure of the $(y,\lambda)$ plane changes drastically. More precisely, dense attracting domains take over, while the fractal regions are mainly confined in the vicinity of the boundaries of the basins of convergence. Looking at panel (b) of Fig. \ref{yl} it becomes more than evident that for $\lambda < \lambda_1$ the Newton-Raphson iterative scheme requires, in general terms, a substantial number of iterations for converging to one of the three equilibrium points. For larger values of $\lambda$ however, the required number of iterations drops significantly.

\begin{figure}[!t]
\centering
\resizebox{\hsize}{!}{\includegraphics{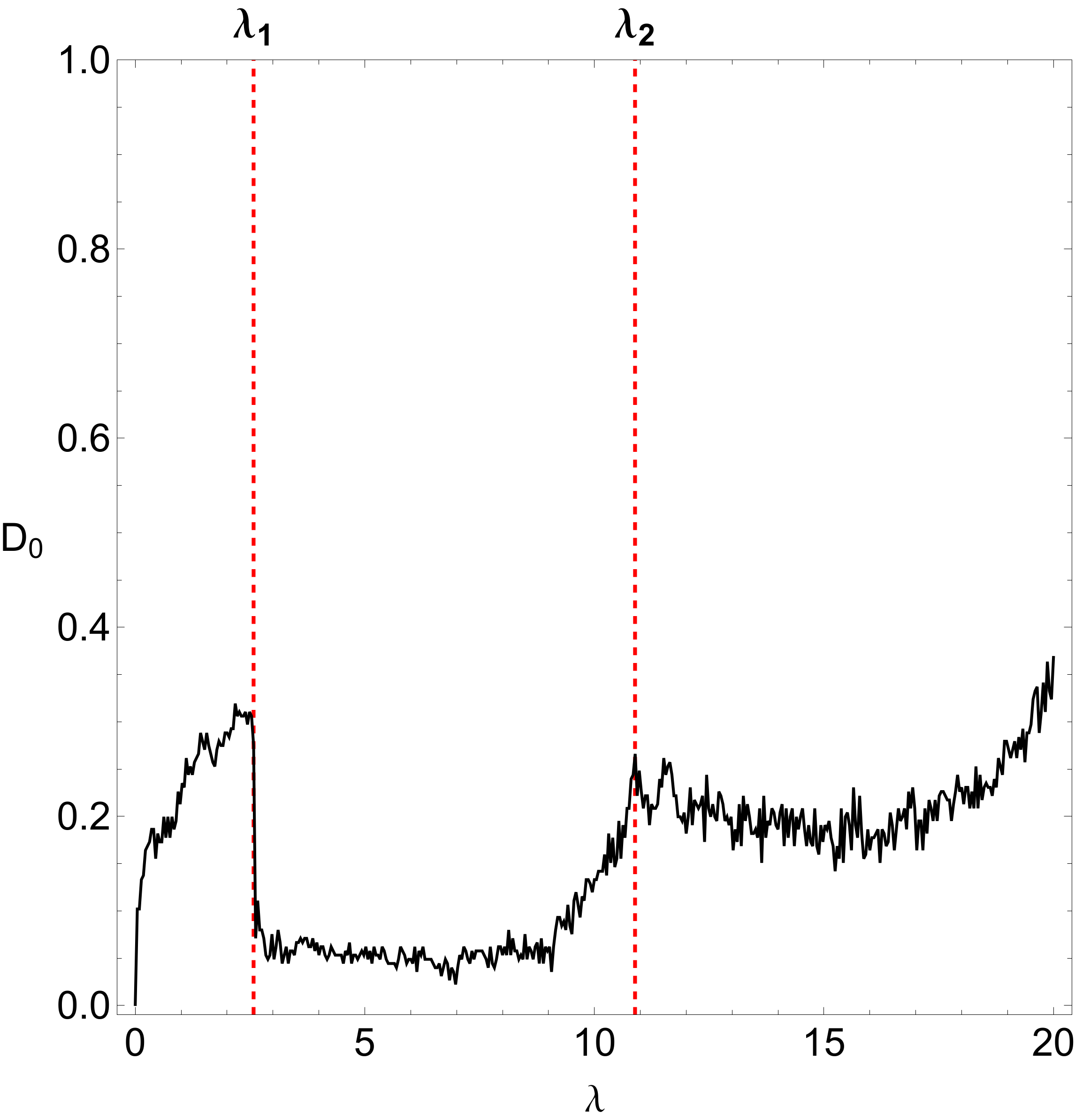}}
\caption{Evolution of the fractal dimension $D_0$ of the $(y,\lambda)$ plane of Fig. \ref{yl} as a function of the ratio of the magnitudes of the magnetic dipoles $\lambda$. $D_0 = 1$ means total fractality, while $D_0 = 0$ implies zero fractality. The red dashed vertical lines indicate the two critical values of parameter $\lambda$, which distinguish between the three cases regarding the total number and of course the type of the equilibrium points.}
\label{frac}
\end{figure}

So far we have discussed the fractality of the several two-dimensional planes only in a qualitative way. More precisely, we seen that the highly fractal areas are those in which we cannot predict from which attractor (equilibrium point) each initial condition is attracted. On the other hand, inside the basins of convergence the degree of fractality is zero and the final state of the initial conditions is well known and of course predictable. At this point we shall provide a quantitative analysis regarding the degree of fractality for the $(x,\lambda)$ plane shown earlier in Fig. \ref{yl}. In order to measure the degree of fractality we have computed the uncertainty dimension \citep{O93} for different values of the ratio of the magnitudes of the magnetic dipoles $\lambda$, thus following the computational method introduced in \cite{AVS01}. Obviously, this degree of fractality is completely independent of the initial conditions we used to compute it.

The evolution of the uncertainty dimension $D_0$ for the $(y,\lambda)$ plane, as a function of the parameter $\lambda$, is shown in Fig. \ref{frac}. The computation of the uncertainty dimension was done for only a ``1D slice'' of initial conditions of Fig. \ref{yl}, and for that reason $D_0 \in (0,1)$. It is interesting to note that in the interval $\lambda \in [\lambda_1, \lambda_2]$ the lowest degree of fractality is observed. Moreover, in general terms we may argue that when $\lambda > 15$ the uncertainty dimension increases with increasing value of the ratio of the magnitudes of the magnetic moments.

\section{Concluding remarks}
\label{conc}

In this this paper we expanded the work initiated in Paper I regarding the Newton-Raphson basins of attraction in the electromagnetic Copenhagen problem where the two primary bodies are magnetic dipoles. In this dynamical system the total number of the libration points strongly depends on the numerical value of the parameter $\lambda$. The basins of convergence leading to the equilibrium points of the dynamical system have been determined with the help of the multivariate version of the Newton-Raphson method. These basins describe how each point on the configuration $(x,y)$ plane is attracted by one of the libration points which act as attractors. Our numerical exploration revealed how the position of the equilibrium points and the structure of the basins of attraction are influenced by the ratio $\lambda$ of the magnetic moments of the primaries. We also found correlations between the attracting domains and the distribution of the corresponding required number of iterations. To our knowledge, this is the first time that such a thorough and systematic numerical investigation, regarding the basins of attraction, takes place in the electromagnetic binary system and this is exactly the novelty as well as the importance of the current work.

For the numerical scan of the sets of the initial conditions on the configuration $(x,y)$ plane, we needed about 4 minutes of CPU time on a Quad-Core i7 2.4 GHz PC, depending of course on the required number of iterations. When an initial condition had converged to one of the attractors with the predefined accuracy the iterative procedure was effectively ended and proceeded to the next available initial condition.

The main results of our numerical research can be summarized as follows:
\begin{enumerate}
  \item In all studied cases, the configuration $(x,y)$ plane is a complicated mixture of basins of attraction and highly fractal regions. These regions are the exact opposite of the basins of attraction and they are completely intertwined with respect to each other (fractal structure). This sensitivity towards slight changes in the initial conditions in the fractal regions implies that it is impossible to predict the final state (attractor).
  \item The several basins of attraction are very intricately interwoven and they appear either as well-defined broad regions, as thin elongated bands, or even as spiraling structures. The structure of the basins of convergence consists of some dense parts that mainly evolve around the respective libration points. The central areas of these dense regions are compact and homogeneous which shows the deterministic aspect of these domains. The fractal domains on the other hand are mainly located in the vicinity of the basin boundaries.
  \item The area of the attracting areas corresponding to collinear equilibrium points $L_2$, $L_3$ as well as to triangular points $L_4$, $L_5$, $L_6$ and $L_7$ is finite. The numerical computations reveal that the area of the basins of convergence corresponding to the central equilibrium point $L_1$ extends to infinity.
  \item Our calculations strongly suggest that non-converging points on the configuration plane exist only when $0 < \lambda < \lambda_1$. Their portion increases as the value of the ratio $\lambda$ tends to $\lambda_1$. On the contrary, for $\lambda \geq \lambda_1$ all initial conditions converge, sooner or later, to one of the attractors of the dynamical system.
  \item The iterative method was found to converge very fast $(0 \leq N < 15)$ for initial conditions around each equilibrium point, fast $(15 \leq N < 25)$ and slow $(25 \leq N < 50)$ for initial conditions that complement the central regions of the very fast convergence, and very slow $(N \geq 50)$ for initial conditions of dispersed points lying either in the vicinity of the basin boundaries, or between the dense regions of the equilibrium points.
  \item We found that the change on the value of the ratio $\lambda$ of the magnetic moments mostly influences the shape and the geometry of the basins of attraction when $0 < \lambda < \lambda_1$. For larger values of $\lambda$ the influence of the same parameter on the attracting regions is milder.
  \item The required number of iterations for obtaining the desired accuracy decreases in the first case ($0 < \lambda < \lambda_1$), then it slightly increases in the second case ($\lambda_1 \leq \lambda \leq \lambda_2$), while throughout the third interval ($\lambda > \lambda_2$) it remains unperturbed.
\end{enumerate}

Taking into account the detailed and novel outcomes of our numerical investigation we may claim that our computational task has been successfully completed. We hope that the present numerical analysis and the corresponding outcomes to be useful in the field of Newton-Raphson basins of convergence in the electromagnetic Copenhagen problem. It is in our future plans to use other iterative schemes, of higher order than that of the Newton-Raphson, and compare the similarities and differences regarding the structure of the basins of attraction.

\section{Future work}
\label{futw}

In this work we investigated in detail the Newton-Raphson basins of convergence associated with the several attractors (equilibrium points) in the electromagnetic Copenhagen problem. Now the natural questions that arise are the following: Is there any connection between the basins of convergence and the orbital dynamics of the dynamical system? Or in other words, can we learn anything interesting regarding the test particle dynamics from the knowledge of the Newton-Raphson dynamics of the equilibrium points of the effective potential?

We suspect that such a connection should exist. This should be true because Eq. (\ref{nrm}) contain both the first and the second order derivatives of the effective potential. The derivatives of the first order determine the equations of motion of the test particle, while the second order derivatives enter the variational equations which are used for determining the stability properties of the test particle's dynamics (i.e., we can use them to calculate the monodromy matrix of the periodic orbits). Therefore, the multivariate Newton-Raphson iterative scheme combine in a (at least at the moment) mysterious way the equations of motion along with the variational equations. In this sense we may claim that the Newton-Raphson method incorporates and combine, in a tricky form, the dynamics of the test particle's orbit together with the corresponding stability properties.

For the Newton-Raphson method the effective potential is used just as a given static function on the configuration space. However, the Newton-Raphson method does not care, that in reality it is an effective potential in a rotating frame. Looking at Eq. (\ref{pot}) we see that the magnetic potential falls off for large distance and then the effective potential converges for large distance to the centrifugal potential. This particular asymptotic behavior of the effective potential might indicate that in reality we are working in a frame rotating with angular frequency $\omega$. This in turn indicates that we should compare with a corresponding particle dynamics in a rotating system. Thus the test particle's dynamics should also include Coriolis forces which are not directed by the effective potential nor included in it. However, indirectly the functional form of the effective potential suggests the need to include these inertial forces into the corresponding particle dynamics.

Taking into consideration all the above-mentioned points it is in our future plans to expand our exploration and try to understand if there is additional hidden, yet very useful, information in the Newton-Raphson basins of convergence in the electromagnetic Copenhagen problem.

\section*{Acknowledgments}

I would like to express my warmest thanks to the two anonymous referees for the careful reading of the manuscript and for all the apt suggestions and comments which allowed us to improve both the quality and the clarity of the paper.

\end{document}